# spike: A Tool to Drizzle HST, JWST, and Roman PSFs for Improved Analyses

**Ava Polzin** *[1]

**1** Department of Astronomy and Astrophysics, The University of Chicago, USA

## Summary

An instrumental point spread function (PSF) describes the distribution of light for a pure point source in an astronomical image due to the optics of the instrument. An accurate PSF is key for deconvolution, point source photometry and source removal. Space-based telescopes can then pose a challenge because their PSFs are influenced by their complex construction, and the myriad of pointings and rotations used to capture deep images. These telescopes also capture the highest resolution images of astronomical sources, resolving stars in even relatively distant galaxies. Proper co-addition of PSFs at a specific source position for space-based imaging is then both critical and challenging. The library described in this work, spike, generates model PSFs and runs them through the same processing pipeline used to derive deep, co-added images, providing correctly co-added and resampled PSFs for images from the Hubble Space Telescope, the James Webb Space Telescope, and the Nancy Grace Roman Space Telescope.

## Statement of need

The PSFs of co-added images are of interest to both ground- and space-based instruments. PSFs can be impacted by different co-addition schemes and may impact the analysis results based those data (Mandelbaum et al., 2023). The cumulative effect of the geometric distortions and offsets in angle and pixel location of space-based data are apparent in the effective PSFs of the co-added and resampled ("drizzled", Fruchter & Hook, 2002) images, making the PSFs modeled on uncombined images insufficient for careful photometric analyses. This is a recognized limitation of existing PSF models, and DrizzlePac recently added functionality to use drizzled pre-computed model PSFs in their native photometric catalog generator (Hack et al., 2021). However, this drizzled PSF uses generic TinyTim (Krist et al., 2011) models, which do not account for the positions of sources on the chip and do not allow the user to set model parameters without overwriting the existing grid of model PSFs. Simultaneously, the resultant co-added PSF is not available to users, and there is not an existing, simple way to generate drizzled PSFs for use in other analyses. Alternatively, an empirical PSF may be created from the data (see, e.g., Liaudat et al., 2023) for a brief review] or unsaturated stars near the object of interest may be selected as a proxy for the PSF. For particularly crowded fields, or those where most stars are saturated, these alternatives pose a problem.

For analyses to be consistent, mock PSFs may be used, but the proper treatment of them involves generating model PSFs and then co-adding and processing them in the same way as the parent images are co-added and processed. Workflows have been created for some instruments (e.g. HST WFC3) and surveys (e.g., Ji et al., 2024), and individual packages have tried to address this in their own data handling, with e.g., grizli (Brammer, 2024) returning drizzled standard empirical PSFs, but, similar to the drizzled PSFs computed by DrizzlePac, such solutions leave little room for users to define the method or specifics of model PSF generation. The code presented here, spike, streamlines the potentially challenging process of realistic PSF generation, taking images and coordinates and directly outputting correctly co-added model PSFs for the Hubble Space Telescope (HST), the James Webb Space Telescope (JWST), and the upcoming Nancy Grace Roman Space Telescope. Model PSFs can be generated using different industry-standard packages, empirical PSFs can be computed from input images, or users can provide PSFs associated with the individual images. `spike` is both easy to use and flexible.

*apolzin@uchicago.edu

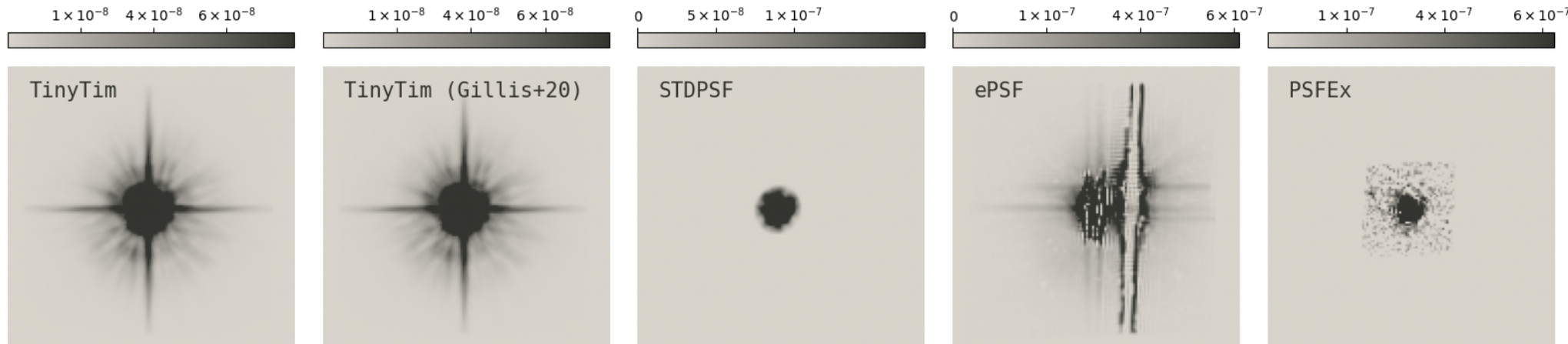

**Figure 1:** Comparison of drizzled PSFs generated for HST images using the default parameters for different methods included in `spike`. All panels use the same ACS/WFC imaging of the COSMOS field in F475W. The stretch of each model's colormap is chosen based on their individual handling of the background, wings, and faint features of the PSF. Note that the ePSF panel (second from right) shows some artifacts; the robustness of the effective PSF method is heavily dependent on the number of stars in the chosen field and may be changed by altering the star detection threshold.

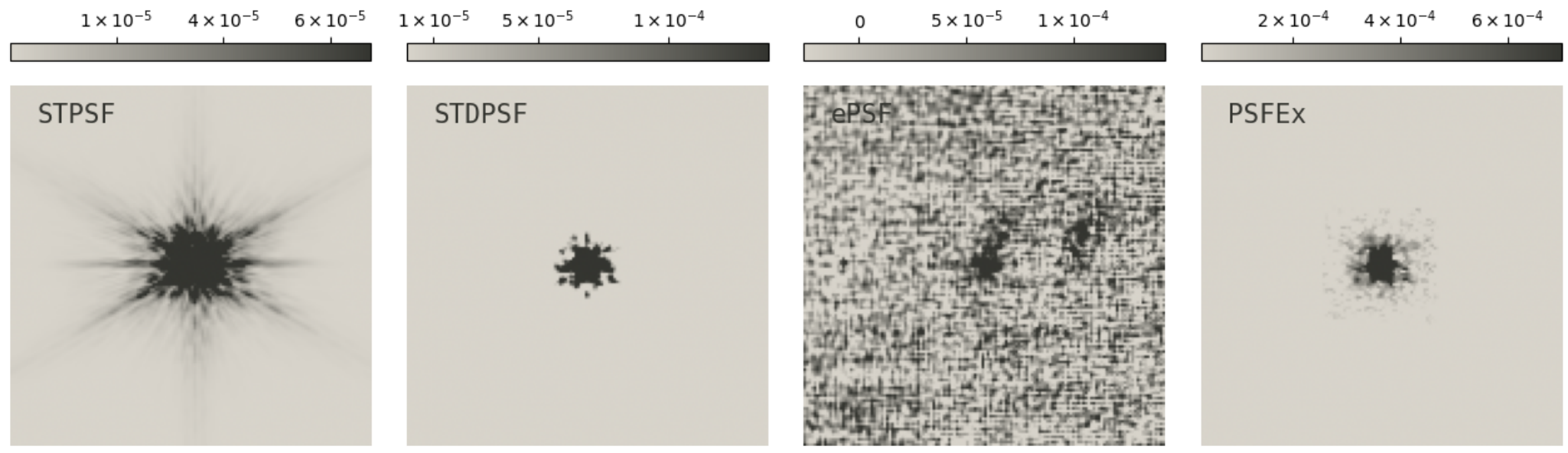

**Figure 2:** Same as Figure 1 for *JWST*/NIRCam imaging in F115W. Note that the ePSF model shown here was generated using a lower detection threshold and a different star selection algorithm due to a paucity of stars in this field.

# Workflow

Given a directory containing reduced and calibrated, but not yet co-added, FITS (Wells et al., 1981) files from HST, JWST, or Roman and the coordinates of an object of interest, detector/chip-specific model PSFs are directly drizzled by `spike`. A model PSF is generated for each unique input image and coordinate combination. Instrument information, including camera, filter, and, if necessary, chip, is automatically read from the header of each FITS file.

`spike` works with both model and empirical PSFs. The built-in PSF generation options are Tiny Tim (Krist et al., 2011) and the Gillis et al. (2020) modification, STPSF (Perrin et al., 2025) – formerly WebbPSF (Perrin et al., 2012, 2014; Perrin et al., 2024), Photutils ePSFs (Anderson & King, 2000; Anderson, 2016; Bradley et al., 2024), PSFEx (Bertin, 2011), and Space Telescope Science Institute's library of empirical STDPSFs for HST and JWST (Anderson, 2016; Libralato et al., 2023, 2024), all of which are included for having the ability to generate a model PSF for an arbitrary detector location. Figure 1 compares output drizzled PSFs from the different methods for HST and Figure 2 compares output drizzled PSFs for JWST.

In most cases, users will only ever interact with the top-level functions spike.psf.hst, spike.psf.jwst, and spike.psf.roman. However, PSF model creation can be accessed directly via the functions in spike.psfgen, and this may be of added value to users on its own, including for use with telescopes/instruments not explicitly mentioned here, as spike smooths over some of the complication of individual tools as an all-in-one means of accessing model PSFs via simple Python functions.

## Preparation for future observatories

There is a preliminary module included in 'spike' for recovering resampled Roman PSFs. spike.psf.roman currently takes as input single detector images, which can be used with STPSF models via spike.psfgen. Since Roman is not yet collecting data, spike.psf.roman is based on the structure of simulated single detector data from e.g., Troxel et al. (2023). When Roman is launched and data become available, spike will be updated to reflect the most current version of the reduction pipeline and the detailed considerations of the real data.

## Managing restrictive dependencies

Since drizzled spike PSFs are intended for use with calibrated and resampled data products from the original pipelines, it is imperative that the processing done on the PSFs is the same as the processing of those data products. Both jwst (Bushouse et al., 2024) and romancal (Desjardins et al., 2023) are complex packages that house the entire JWST and Roman pipelines. As such, they have complicated functionality that relies on modules with more stringent installation requirements. Using jwst and romancal out of the box places strict limitations on the allowed operating systems. To address this, there are stripped down versions of the necessary "tweak" and resample scripts included with spike as spike.jwstcal, spike.romancal, spike.stcal, and spike.stpipe. Each module is only subtly changed from jwst, romancal, and their underlying STCAL and stpipe to avoid unnecessary dependencies that restrict installation.

## Software and Packages Used

- ASDF (Greenfield et al., 2015; Graham et al., 2024)
- Astropy (Astropy Collaboration et al., 2013, 2018, 2022)
- CRDS (Greenfield & Miller, 2016)
- drizzle (Simon et al., 2024a)
- drizzlepac (Fruchter & et al., 2010; Hoffmann et al., 2021)
- GWCS (Dencheva et al., 2024)
- jsonschema (Berman et al., 2024)
- jwst (Bushouse et al., 2024)
- Matplotlib (Hunter, 2007)
- NumPy (Harris et al., 2020)
- Photutils (Bradley et al., 2024)
- PSFEx (Bertin, 2011)
- psutil
- PyYAML

- roman_datamodels (Huwe et al., 2025)
- romancal (Desjardins et al., 2023)
- SciPy (Virtanen et al., 2020)
- SExtractor (Bertin & Arnouts, 1996)
- spherical_geometry (Simon et al., 2024b)
- STCAL
- stdatamodels
- stpipe
- Tiny Tim (Krist et al., 2011) – including the option to use the Gillis et al. (2020) parameters
- tweakwcs (Cara et al., 2024)
- STPSF (Perrin et al., 2025), formerly WebbPSF (Perrin et al., 2012, 2014; Perrin et al., 2024)

## Acknowledgments

AP is supported by the Quad Fellowship administered by IIE and thanks Hsiao-Wen Chen and Juan Guerra for their comments. AP also thanks the reviewers for their suggestions, which improved the manuscript and codebase. This work makes use of observations made with the NASA/ESA Hubble Space Telescope and the NASA/ESA/CSA James Webb Space Telescope obtained from the Mikulski Archive for Space Telescopes at the Space Telescope Science Institute, which is operated by the Association of Universities for Research in Astronomy, Inc., under NASA contracts NAS 5–26555 for HST and NAS 5-03127 for JWST.